\documentclass[12pt]{article}
\usepackage{amsmath,amssymb,amsthm,mathrsfs,url,hyperref}

\title{Empirically Equivalent Distributions in Ontological Models of Quantum Mechanics}

\author{
Roderich Tumulka\footnote{Fachbereich Mathematik, Eberhard-Karls-Universit\"at, 
	Auf der Morgenstelle 10, 72070 T\"ubingen, Germany. 
	E-mail: roderich.tumulka@uni-tuebingen.de}
}

\date{May 9, 2022}

\newcommand{\Hilbert}{\mathscr{H}}
\newcommand{\be}{\begin{equation}}
\newcommand{\ee}{\end{equation}}

\newcommand{\RRR}{\mathbb{R}}
\newcommand{\SSS}{\mathbb{S}}
\newcommand{\CCC}{\mathbb{C}}

\newcommand{\PPP}{\mathbb{P}}

\newcommand{\cC}{\mathcal{C}}
\newcommand{\cEXP}{\mathcal{EXP}}
\newcommand{\cPOVM}{\mathcal{POVM}}
\newcommand{\cDM}{\mathcal{DM}}
\newcommand{\cPM}{\mathcal{PM}}
\newcommand{\sE}{\mathscr{E}}
\newcommand{\sF}{\mathscr{F}}
\newcommand{\pr}[1]{|#1\rangle\langle #1|}
\DeclareMathOperator{\tr}{tr}
\theoremstyle{theorem}
\newtheorem{thm}{Theorem}

\begin{document}
\maketitle
\begin{abstract}
We consider ontological models of a quantum system, assuming that not all probability distributions over the space $\Lambda$ of ontic states are preparable, only those belonging to a certain set $\cC$. We assume further that every POVM with a finite value space can be measured and that for every density matrix there exists a distribution in $\cC$ whose outcome statistics is given by the density matrix. We show that this mapping from $\cC$ to the set of density matrices must be many-to-one, that is, that there must be empirically indistinguishable distributions in $\cC$. This shows that there must be limitations to knowledge in the sense of facts in nature that cannot be discovered empirically.

\medskip

\end{abstract}

\section{Introduction}

This paper describes a previously unpublished result from my 1998 Diplom thesis \cite{Tum98}. 

It is well known that two different ensembles $\mu_a\neq \mu_b$ of wave functions can have the same density matrix. Suppose that an observer Alice chooses either $x=a$ or $x=b$, prepares a large number of quantum systems with $\mu_x$-distributed wave functions, hands the systems over to another observer Bob and challenges him to find out which ensemble she chose. Since the distribution of outcomes of any experiment depends only on the density matrix, Bob cannot distinguish between the two possibilities. This fact shows, for any interpretation of quantum mechanics according to which nature knows the wave function (at least up to a phase), that there must be facts in reality, known to nature, that we cannot discover empirically (such as whether the ensemble of systems is actually $\mu_a$- or $\mu_b$-distributed). We call this conclusion a \emph{limitation to knowledge}.

The question arises whether \emph{every} empirically adequate interpretation of quantum mechanics must entail limitations to knowledge. Here, we present a mathematical theorem intended to show that the answer is Yes. 

The same conclusion can also be drawn from the Pusey--Barrett--Rudolph (PBR) theorem \cite{PBR,Lei14} (2011), but the proof presented here is independent of the PBR theorem. As far as I know, this proof, first written in 1998, may have been the earliest proof that \emph{every} quantum theory poses limitations to knowledge.

In order to allow for a full generality of interpretations or theories of quantum mechanics, we let the theory decide what the \emph{ontic states} (the possible physically real situations) of a quantum system are; we denote the set of all ontic states $\lambda$ by $\Lambda$. The theory may assert that it is impossible to prepare a system in a particular ontic state $\lambda$ of our choice, or in a particular probability distribution over $\Lambda$ of our choice; instead, the theory may assert that only probability distributions from a certain class $\cC$ can be prepared. (For example, in Bohmian mechanics of $N$ particles, where the ontic states are pairs $(Q,\Psi)\in\RRR^{3N}\times L^2(\RRR^{3N},\CCC^k)$ of configuration and wave function, we can only prepare distributions with density $\varrho(q,\psi)=|\psi(q)|^2\,\delta(\psi-\phi)$ for some wave function $\phi$, or mixtures thereof.) For any experiment that we can carry out on the system, the theory must provide the probability distribution of the outcome for every $\lambda$. A theory in this sense is sometimes called an \emph{ontological model} \cite{Lei14}. We show that if the model is empirically adequate (i.e., if the theory's empirical predictions agree with the standard rules of quantum mechanics), then in $\cC$ there must be distinct distributions $\varrho_a\neq \varrho_b$ that are empirically indistinguishable.

The PBR theorem shows that in any empirically adequate ontological model under a reasonable further assumption, the ontic state $\lambda$ contains full information about the wave function $\psi$ up to phase. However, before or without the PBR theorem, we would have to take into account the possibility that the same $\lambda$ could arise from different preparation procedures corresponding to different $\psi$'s. There are examples of such ``$\psi$-epistemic'' ontological models (which violate the further assumption of PBR): Kochen and Specker \cite{KS67} described an empirically adequate ontological model (with $\Lambda$ the unit sphere in $\RRR^3$) of a single spin, Spekkens \cite{Spe04} a model (with $\Lambda$ having 4 elements) of a single spin that accounts for quantum measurements of the 3 Pauli matrices (see also \cite[Sec.s~2 and 4.3]{Lei14}). Finally, with the possibility of $\psi$-epistemic models, we would also have to take into account the possibility that two distinct distributions over $\psi$'s might lead to the same distribution over $\lambda$'s, so that the existence of empirically equivalent distributions over $\psi$'s need not involve a limitation to knowledge, as it leaves open whether distinct distributions of $\lambda$'s can always be distinguished empirically. This possibility is excluded by our theorem.

\section{Result}

For any measurable space $(\Omega,\sF)$, let $\cPM(\Omega)$ denote the set of all probability measures on $\Omega$.
We consider a quantum system with Hilbert space $\Hilbert$ with $\dim\Hilbert\geq 2$. Of an \emph{ontological model}, we demand the following. We are given a measurable space $(\Lambda,\sF)$ whose elements $\lambda$ are called the ontic states. Let $\cPOVM$ denote the set of all POVMs acting on $\Hilbert$ with finite value space $\subset \RRR$ and $\cDM$ the set of all density matrices in $\Hilbert$. Let $\cEXP$ be an index set called the set of experiments on this system. Since every quantum experiment is associated with a POVM, we assume that this association is expressed through a mapping $E:\cEXP\to\cPOVM$, and we assume that $E$ is surjective (so every finite POVM can be ``measured''). For any ontic state $\lambda\in\Lambda$ and experiment $\sE\in\cEXP$, let $\PPP_{\lambda,\sE}$ denote the probability distribution of the outcome, given the system is in $\lambda$; so, $\PPP$ is a mapping $\Lambda\times \cEXP\to \cPM(\RRR)$ (using the Borel $\sigma$-algebra of $\RRR$). We assume that the set $\cC\subseteq \cPM(\Lambda)$ of preparable distributions is convex (because, if we have two preparation procedures, we might choose one of them randomly). We assume further that for every $\varrho\in\cC$, there is a density matrix $D_\varrho$ on $\Hilbert$ such that, for every experiment $\sE\in\cEXP$, the distribution of outcomes agrees with the quantum mechanical prediction for $D_\varrho$,
\be\label{agree}
\int_\Lambda \varrho(d\lambda) \, \PPP_{\lambda,\sE}(B) = \tr \Bigl( D_\varrho \, E_{\sE}(B) \Bigr)
\ee
for every $B\in\sF$.
We assume further that $D:\cC\to\cDM$ is surjective (so every quantum state can be accounted for by the model).
These assumptions together form our definition of an ontological model.

\begin{thm}
The mapping $D:\cC\to\cDM$ cannot be injective.
\end{thm}

\section{Proof}

It suffices to consider the case $\dim\Hilbert=2$; for every other Hilbert space, the result follows by restriction on a 2d subspace. We assume that $D$ is injective and will derive a contradiction. 

First, we show that $D$ is convex linear. If $\varrho_a, \varrho_b\in \cC$ and $s\in[0,1]$, then $\varrho:= s\varrho_a +(1-s) \varrho_b$ lies in $\cC$ by assumption, and
\begin{subequations}
\begin{align}
\tr\Bigl( D_{\!\varrho} \, E_{\sE}(B)  \Bigr)
&\stackrel{\eqref{agree}}{=}
\int_\Lambda\varrho(d\lambda) \, \PPP_{\lambda\sE}(B) \\
&=s\int_\Lambda\varrho_a(d\lambda) \, \PPP_{\lambda\sE}(B) + (1-s) \int_\Lambda\varrho_b(d\lambda) \, \PPP_{\lambda\sE}(B) \\
&\stackrel{\eqref{agree}}{=} 
s\tr\Bigl( D_{\varrho_a} \, E_{\sE}(B) \Bigr)
+(1-s)\tr\Bigl( D_{\varrho_b} \, E_{\sE}(B) \Bigr)\\
&=\tr\Bigl( (sD_{\varrho_a}+(1-s)D_{\varrho_b}) E_{\sE}(B) \Bigr) \,.
\end{align} 
\end{subequations}
Since $E$ is surjective, every positive operator $\leq I$ can occur as $E(B)$. The relation $\tr(D_1 \,E(B)) = \tr(D_2 \, E(B))$ can only hold for all $0\leq E(B)\leq I$ (which includes all 1d projections) if $D_1=D_2$; thus, $D_{\varrho}=sD_{\varrho_a}+(1-s)D_{\varrho_b}$.

Second, since $D$ is bijective, it has an inverse $D^{-1}:\cDM\to\cC$. The inverse of any bijective convex linear mapping $D$ between convex sets is convex linear because if $W_a,W_b\in \cDM$ and $W:= sW_a+(1-s)W_b$ with $s\in[0,1]$, set $\varrho_a:=D^{-1}(W_a)$, $\varrho_b:= D^{-1}(W_b)$, $\varrho:=s\varrho_a +(1-s)\varrho_b$, and observe that $D(\varrho) = sD(\varrho_a)+(1-s)D(\varrho_b)= s W_a + (1-s) W_b = W$, so $\varrho=D^{-1}(W)$.

Next, define the measure
\be
\varrho_1:= 2 \, D^{-1}(\tfrac12 I)
\ee
(the subscript 1 indicates that it is the ``whole'' measure, or the one corresponding to the identity operator). For every 1d subspace $g$ of $\Hilbert$, define
\be
\varrho_{g} := D^{-1}(P_g)
\ee
with $P_g$ the projection to $g$, $P_g=\pr{\psi}$ for $\psi\in\SSS(g)$. From $P_g+P_{g^\perp}=I$ it follows by convex linearity of $D^{-1}$ that
\be\label{rhogrho1}
\varrho_g + \varrho_{g^\perp} = \varrho_1\,.
\ee
Let $h$ be a 1d subspace and consider the POVM with value space $\{0,1\}$ consisting of the operators $P_{h^\perp}$ and $P_h$. Since $E$ is surjective, we can choose for every $h$ an $\sE_h\in\cEXP$ so that $E_{\sE_h}(0)=P_{h^\perp}$ and $E_{\sE_h}(1)=P_h$. By \eqref{agree},
\be\label{rhog}
\int_\Lambda \varrho_g(d\lambda) \, \PPP_{\lambda,\sE_h} = \tr(P_g P_h) \delta_1 + \tr(P_g P_{h^\perp}) \delta_0\,,
\ee
where $\delta_x$ means the measure on $\RRR$ with weight 1 in the point $x$ and 0 everywhere else (i.e., with ``density'' $\delta(\cdot - x)$). Define the subset $\Lambda_h \subseteq \Lambda$ by
\be
\Lambda_h := \bigl\{ \lambda\in\Lambda: \PPP_{\lambda,\sE_h}=\delta_1 \bigr\}\,.
\ee
Since 0 and 1 are the only possible outcomes of $\sE_h$, we have that $\PPP_{\lambda,\sE_h}$ is of the form $p_{\lambda,\sE_h}\delta_1+(1-p_{\lambda,\sE_h})\delta_0$ with $p_{\lambda,\sE_h}\in[0,1]$ for $\varrho$-almost every $\lambda$ for every $\varrho\in\cC$ (as well as for $\varrho=\varrho_1$). In particular, $\Lambda_h=\{\lambda\in\Lambda: p_{\lambda,\sE_h} =1\}$. By \eqref{rhog},
\be\label{rhogphtr}
\int_\Lambda \varrho_g(d\lambda) \, p_{\lambda,\sE_h} = \tr(P_g P_h)\,.
\ee
For $h=g$ we obtain that 
\be
\int_\Lambda \varrho_g(d\lambda) \, p_{\lambda,\sE_g} = 1,
\ee
but since $p_{\lambda,\sE_g}\leq 1$, this can only happen if $p_{\lambda,\sE_g}=1$ for $\varrho_g$-almost every $\lambda$; thus,
\be\label{rhog1}
\varrho_g(\Lambda_g)=1\,,
\ee
i.e., the probability measure $\varrho_g$ is concentrated in $\Lambda_g$.

For $h=g^\perp$ we obtain from \eqref{rhog} that 
\be
\int_\Lambda \varrho_g(d\lambda) \, p_{\lambda,\sE_{g^\perp}}=0\,,
\ee
and since $p_{\lambda,\sE_{g^\perp}}\geq 0$, this can only happen if $p_{\lambda,\sE_{g^\perp}}=0$ for $\varrho_g$-almost every $\lambda$; in particular,
\be\label{rhog0}
\varrho_g(\Lambda_{g^\perp})=0\,,
\ee
but also
\be\label{rhogbetween}
\varrho_g(\{\lambda\in\Lambda: 0\neq p_{\lambda,\sE_{g^\perp}} \neq 1\})=0\,.
\ee
Since $g$ was arbitrary, we also have that
\begin{subequations}
\begin{align}
\varrho_{g^\perp}(\Lambda_{g^\perp})&=1\,, \label{rhogperp1}\\
\varrho_{g^\perp}(\Lambda_g)&=0\,. \label{rhogperp0}
\end{align}
\end{subequations}
By \eqref{rhogrho1}, for any set $B\in\sF$ in $\Lambda$,
\begin{subequations}
\begin{align}
\varrho_1(B)
&=\varrho_g(B) + \varrho_{g^\perp}(B)\\
&\stackrel{\eqref{rhog1},\eqref{rhogperp1}}{=}\varrho_g(B\cap \Lambda_g) + \varrho_{g^\perp}(B\cap \Lambda_{g^\perp})\,. \label{rho1ggperp}
\end{align}
\end{subequations}
In particular,
\begin{subequations}
\begin{align}
\varrho_1(\Lambda_g\cap \Lambda_{g^\perp}) 
&=\varrho_g(\Lambda_g\cap \Lambda_{g^\perp}) 
+\varrho_{g^\perp}(\Lambda_g\cap \Lambda_{g^\perp}) \\
&\leq \varrho_g(\Lambda_{g^\perp}) 
+\varrho_{g^\perp}(\Lambda_g)
\stackrel{\eqref{rhog0},\eqref{rhogperp0}}{=}0
\end{align}
\end{subequations}
and
\begin{subequations}
\begin{align}
\varrho_g(B)
&\stackrel{\eqref{rhog1}}{=}\varrho_g(B\cap\Lambda_g)\\
&\stackrel{\eqref{rhogperp0}}{=} \varrho_g(B\cap\Lambda_g) + \underbrace{\varrho_{g^\perp}(B\cap \Lambda_g)}_{=0 \text{ by \eqref{rhogperp0}}} \\
&\stackrel{\eqref{rhogrho1}}{=}\varrho_1(B\cap \Lambda_g)\,. \label{rhogrho1Lambdag}
\end{align}
\end{subequations}
So we have expressed $\varrho_g$ in terms of $\varrho_1$ and $\Lambda_g$. It follows from \eqref{rho1ggperp} and \eqref{rhogbetween} for $g=h^\perp$ that
\be\label{rho1between}
\varrho_1(\{\lambda\in\Lambda: 0\neq p_{\lambda,\sE_h} \neq 1\})=0 \,.
\ee
We obtain further that, for any 1d subspaces $g,h$,
\begin{subequations}
\begin{align}
\varrho_1(\Lambda_g \cap \Lambda_h)
&= \int_\Lambda \varrho_1(d\lambda) \, 1_{\Lambda_g}(\lambda) \: 1_{\Lambda_h}(\lambda)\\
&\stackrel{\eqref{rho1between}}{=}
\int_\Lambda \varrho_1(d\lambda) \, 1_{\Lambda_g}(\lambda) \: p_{\lambda,\sE_h}\\
&\stackrel{\eqref{rhogrho1Lambdag}}{=}
\int_\Lambda \varrho_g(d\lambda) \, p_{\lambda,\sE_h}\\
&\stackrel{\eqref{rhogphtr}}{=} 
\tr(P_g \, P_h) \,.\label{rho1LambdagLambdah}
\end{align}
\end{subequations}

The remainder of the proof will show that there are no sets $\Lambda_g$ in any measure space $(\Lambda,\sF,\varrho_1)$ with $\varrho_1(\Lambda)=2$ that make the relation \eqref{rho1LambdagLambdah} true. If there were, let $\lambda$ be a random point in $\Lambda$ with distribution $\tfrac{1}{2}\varrho_1$ and define the $\{0,1\}$-valued random variables $X_g=1_{\Lambda_g}(\lambda)$. Since $\varrho_1(\Lambda_g) = \varrho_1(\Lambda_g\cap \Lambda_g) = \tr( P_g \, P_g) = \tr(P_g) =1$, we have for any $g,h$ that
\begin{subequations}
\begin{align}
\PPP(X_g=1,X_h=1) &= \tfrac12 \tr(P_g \, P_h)\\
\PPP(X_g=0,X_h=1) &= \tfrac12-\tfrac12 \tr(P_g \, P_h)\\
\PPP(X_g=1,X_h=0) &= \tfrac12-\tfrac12 \tr(P_g \, P_h)\\
\PPP(X_g=0,X_h=0) &= \tfrac12 \tr(P_g \, P_h)\,.
\end{align}
\end{subequations}
Now consider for $g$ and $h$ the three subspaces
\be
g_1=\CCC\begin{pmatrix}1\\0\end{pmatrix}\,,~~~
g_2=\CCC\begin{pmatrix}1/2\\ \sqrt{3}/2\end{pmatrix}\,,~~~
g_3=\CCC\begin{pmatrix}1/2\\-\sqrt{3}/2\end{pmatrix}\,.
\ee
Since for any two of these, $\tr(P_g \, P_h)=1/4$, any two of the three random variables $X_{g_i}$ would have to have joint distribution
\be
\begin{array}{r|c|c|}
&0&1\\\hline
0&1/8&3/8\\\hline
1&3/8&1/8 \\\hline
\end{array}~.
\ee
If $p_{ijk}$ denotes the probability that $X_{g_1}=i, X_{g_2}=j, X_{g_3}=k$, then
\begin{subequations}
\begin{align}
p_{000}+p_{001}&=1/8 \label{psuma}\\
p_{001}+p_{011}&=3/8 \label{psumb}\\
p_{011}+p_{111}&=1/8 \label{psumc}\,,
\end{align}
\end{subequations}
and \eqref{psuma}$-$\eqref{psumb}+\eqref{psumc} yields $p_{000}+p_{111}=-1/8$, which is impossible.
%
%

\section{Conclusion}

Theorem 1 shows that any theory of quantum reality must entail limitations to knowledge, specifically that there are facts in nature that observers cannot determine empirically. 
Limitations to knowledge run against the positivistic idea that physical theories should involve only observable quantities. But in quantum mechanics, limitations to knowledge are a fact.

Another proof showing that limitations to knowledge arise in every theory of quantum mechanics can be obtained from the Pusey--Barrett--Rudolph theorem \cite{PBR}, which asserts that every ontological model of quantum mechanics satisfying a reasonable assumption must be $\psi$-ontic, together with the fact that $\psi$ cannot be measured (or the fact that different ensembles of $\psi$'s can have the same density matrix).

\end{document}